\documentclass[%
 reprint,
 amsmath,amssymb,
 aps,
]{revtex4-2}
\usepackage{graphicx}
\usepackage{bm}
\usepackage{array}[=2016-10-06]
\usepackage{xcolor}
\usepackage{subfig}

\begin{document}

\preprint{APS/123-QED}

\title{Quantum Remote Implementation of Hybrid Operations on Hyperstates Using Hyperentangled States}
\author{Satish Kumar$^1$}
\altaffiliation[]{mr.satishseth@gmail.com}
\author{Anirban Pathak$^2$}
 \email{anirban.pathak@gmail.com}
 \affiliation{$^1$Department of Physical Sciences, Indian Institute of Science Education \& Research (IISER) Mohali, Sector 81, SAS Nagar, Manauli PO, Punjab-140306, India.}
 
\affiliation{$^2$Department of Physics and Materials Science and Engineering, Jaypee Institute of Information Technology, A-10, Sector-62, Noida, Uttar Pradesh-201309, India.}


\date{\today}

\begin{abstract}
Quantum remote control, also known as quantum remote implementation of an operator (QRIO), enables the remote manipulation of an arbitrary quantum state by implementing a desired quantum operation at a distant location. Significant progress has recently been made in developing QRIO protocols and their variants. Most existing schemes employ hyperentangled states where entanglement is shared across multiple degrees of freedom (DOFs). However, these protocols typically exploit only one degree of freedom at a time. In this work, we propose a QRIO protocol that simultaneously utilizes the polarization and spatial DOFs of a two-qubit hyperentangled state to remotely implement an arbitrary hybrid operator on an unknown single-photon two-qubit hyperstate. The shared hyperentangled resource is realized using the polarization and spatial modes of photons, while the protocol is constructed using linear optical elements and cross-Kerr nonlinear interactions to facilitate effective photon-photon coupling. Furthermore, the effects of measurement errors arising from finite coherent state distinguishability and coherent state dissipation are analyzed and the corresponding success probability of the protocol is evaluated. The results demonstrate that an appropriate choice of the cross-Kerr phase shift and coherent state amplitude significantly enhances the protocol performance, making the proposed scheme a promising candidate for hybrid quantum communication and distributed quantum information processing.
\end{abstract}

\maketitle


\section{Introduction\label{Sec:intro}}
Quantum entanglement is one of the most remarkable features of quantum mechanics and constitutes a fundamental resource for a wide range of quantum information processing tasks (\cite{zeilinger2023dance,pathak2013elements,chandru2023random, dutta2025simultaneous} and references therein). Unlike classical correlations, quantum entanglement gives rise to correlations that cannot, in general, be explained within the framework of classical physics. This counterintuitive phenomenon which Einstein famously referred to as ``spooky action at a distance,'' has not only deepened our understanding of quantum theory but has also laid the foundation for several revolutionary quantum technologies. Notably, protocols such as quantum teleportation \cite{Teleportation1993} and superdense coding \cite{DenseCoding1992} exploit shared entanglement to accomplish information processing tasks that have no classical analogue.

Quantum teleportation is one of the most celebrated achievements of quantum information science introduced by Bennett \emph{et al.} \cite{Teleportation1993} that enables the faithful transfer of an unknown quantum state from one distant party to another without physically transmitting the quantum system itself. Instead, the protocol exploits a shared entangled state together with local operations and classical communication (LOCC). The success of quantum teleportation has inspired the development of several protocols for different tasks related to the remote manipulation of quantum states. One such important task is the quantum remote implementation of an operator (QRIO), also referred to as quantum remote control, where the objective is to implement a quantum operation on a distant quantum state without physically transferring either the operator or the quantum state \cite{QRIO01_2001}.

The concept of quantum remote control was first introduced by Huelga \emph{et al.} \cite{QRIO01_2001}, who demonstrated the remote implementation of restricted classes of unitary operations. Subsequently, several protocols were proposed for remotely implementing partially known and arbitrary quantum operations using different classes of entangled quantum channels \cite{RIPUO01_2006,RIHO01_2007,PhyScr24QRIO}. These studies were later extended to propose schemes for controlled, bidirectional, cyclic, joint and multi-party remote implementation of different type of operators, significantly enriching the family of QRIO protocols \cite{CRIPUO01_2007,CRIPUO02_2008,CRIO01_2022,BCCRIPUO_2022,CCRIPUO_2019,CRIO02_2023,CRIPUO_2022,CJRIO01_2024,BCRIO_2025,CBRIO_2026,BRIHO_BRIPUO_2026,CJRIO01_2024}. More recently, the QRIO framework has been further extended to probabilistic joint remote implementation protocols employing nonmaximally hyperentangled states~\cite{wang2026PQRIO}. Besides their fundamental importance, these protocols have potential applications in distributed quantum computation, delegated quantum processing, quantum networking and cloud-based quantum information processing, where quantum operations must often be executed at remote network nodes.

In parallel with the development of QRIO protocols, there has been growing interest in exploiting multiple degrees of freedom (DOFs) of a single-photon for quantum information processing \cite{chandru2025gate}. Photons possess several independent DOFs such as polarization, spatial path, orbital angular momentum, time-bin and frequency, allowing more than one logical qubit to be encoded within the same physical carrier (photon). Such hybrid encoding increases the information carrying capacity while reducing the number of physical photons required for implementing quantum communication tasks. Consequently, quantum information encoded in multiple DOFs has found widespread applications in high-capacity quantum communication, quantum computation and quantum networking \cite{chandru2025gate, kwiat1997production}.

Recently, hyperentangled states \cite{barreiro2005generation} have also been employed as quantum channels for the remote implementation of quantum operations \cite{CRIO01_2022,JRIO01_2022,CJRIO01_2024,QRIO02_2024,CBRIO_2026}. However, despite the availability of entanglement in multiple DOFs, existing QRIO protocols generally utilize only one degree of freedom at a time, while the remaining DOFs serve merely as auxiliary resources. Consequently, the full potential of hyperentanglement for quantum information processing has yet to be fully exploited. Motivated by this observation, it is natural to ask whether multiple DOFs can be exploited simultaneously to remotely implement a hybrid quantum operations acting on a single-photon two-qubit hyperstate. Addressing this question constitutes the principal motivation of the present work.

In this work, we investigate the remote implementation of hybrid quantum operations by simultaneously exploiting multiple DOFs of a hyperentangled state. Specifically, we develop a QRIO protocol for implementing an arbitrary hybrid operator on an unknown single-photon two-qubit hyperstate encoded in the polarization and spatial DOFs. In contrast to the existing approaches, the proposed scheme fully utilizes the entanglement available in both DOFs of the shared polarization-path Bell state simultaneously, thereby enhancing the resource utilization for hybrid quantum information processing. We further examine the performance of the protocol under realistic experimental conditions by incorporating the effects of finite coherent state distinguishability and coherent state dissipation and identify the parameter regime that enables reliable remote implementation with a high success probability.

The rest of the paper is organized as follows. In Sec.~\ref{Sec:qrio_protocol}, we present the proposed QRIO protocol and describe its theoretical framework. Section~\ref{Sec:results} investigates the effects of measurement errors and coherent state dissipation on the protocol performance. Finally, Sec.~\ref{Sec:conclusion} summarizing the main results and discusses possible future research directions.

\section{QRIO on a hyperstate\label{Sec:qrio_protocol}}
\begin{figure*}
    \centering
    \includegraphics[width=\textwidth]{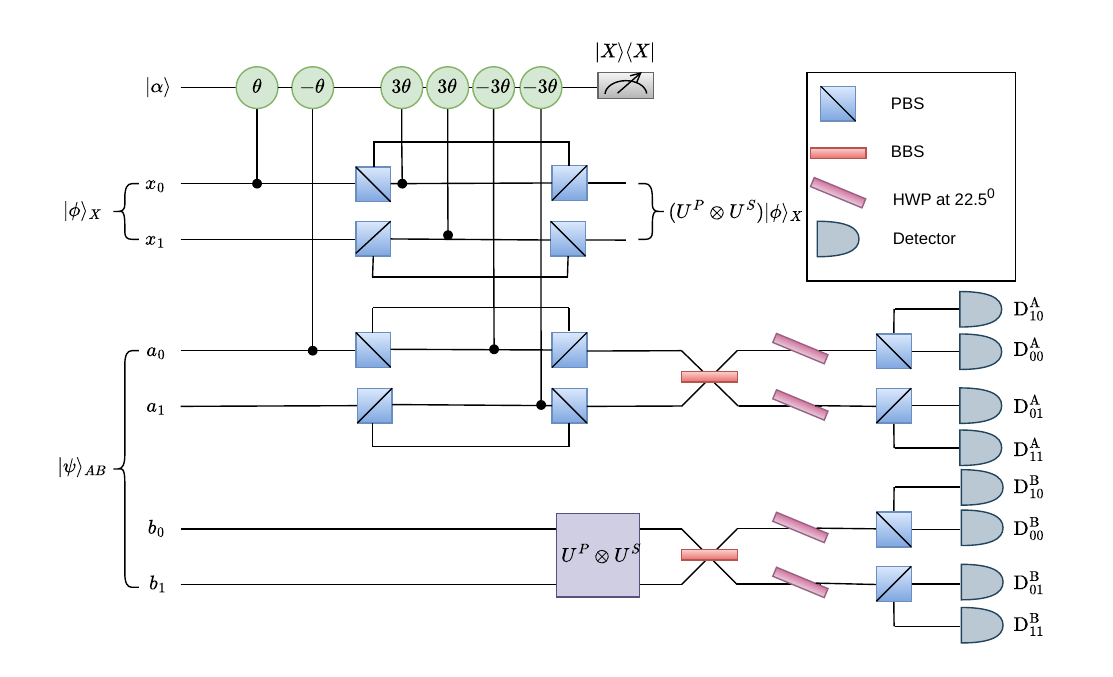}
    \caption{(Color Online) The pictorial representation of the QRIO protocol using a hyperentangled state, where both the spatial and polarization DOFs are utilized simultaneously.}
    \label{Fig:QRIO_Pic}
\end{figure*}

Let Alice and Bob be two distant parties. Alice possesses an arbitrary unknown single-photon two-qubit hybrid state (or hyperstate) of the form
\begin{equation}
|\phi\rangle_{X} =\mu|Hx_0\rangle+\eta|Vx_0\rangle+\gamma|Hx_1\rangle+\delta|Vx_1\rangle\\
\end{equation}
where $|\mu|^2+|\eta|^2+|\gamma|^2+|\delta|^2=1$ and $X$ labels the photon carrying the unknown single-photon two-qubit hyperstate. Here, the first qubit is encoded in the polarization DOF, whereas the second qubit is encoded in the spatial DOF of the photon. Here, $H$ and $V$ denote the horizontal and vertical polarization states, respectively, while $x_0$ and $x_1$ represent the two spatial modes of the photon. Bob’s task is to remotely implement an arbitrary hybrid operator $U=U^P\otimes U^S$ on the unknown hyperstate such that it transforms into $(U^P\otimes U^S)|\phi\rangle_X$, where the superscript $P$ and $S$ refers to the polarization and spatial DOFs, respectively. To accomplish this task, Alice and Bob initially share a polarization-path hyperentangled state given by
\begin{equation}
\begin{split}
|\psi\rangle_{AB}&=|\psi^{P}\rangle_{AB}\otimes|\psi^{S}\rangle_{AB}\\
& =\frac{1}{2}(|Ha_0\rangle_A|Hb_0\rangle+|Va_0\rangle_A|Vb_0\rangle+|Ha_1\rangle_A|Hb_1\rangle\\
&\quad+|Va_1\rangle_A|Vb_1\rangle)_{AB},
\end{split}
\end{equation}
where $a_0$ \& $a_1$ ($b_0$ \& $b_1$) are two paths of photon $A$ ($B$) with Alice (Bob). We will omit the normalization coefficient $1/{\sqrt{2}}$ to simplify the mathematical formulation. \\
The initial three photon joint state involved in the scheme can be written as $|\Phi_0\rangle_{XAB}=|\phi\rangle_{X}\otimes|\psi\rangle_{AB}.$ To initiate the protocol, Alice first entangles the unknown hyperstate with the shared hyperentangled state through the cross-Kerr nonlinear interaction (cK-interaction). The cross-Kerr effect is a third-order nonlinear optical process in which the phase of one optical field is shifted by an amount proportional to the intensity of another interacting field \cite{feizpour2015cK01}. In the present protocol, a coherent state (CS) acts as a probe field and acquires different phase shifts depending on the quantum state of the interacting photons. As a result, the information encoded in the photonic polarization and spatial degrees of freedom is transferred to the phase of the coherent state without destroying the photons. The induced phase shifts are subsequently discriminated through homodyne measurement, thereby generating the required entanglement between the unknown state and the shared hyperentangled resource. To this end, Alice prepares an auxiliary CS $|\alpha\rangle$ and allows it to interact with one of the spatial modes of photons $X$ and $A$, namely $x_0$ and $a_0$ with interaction parameter $\theta$ and $-\theta$, respectively. Subsequently, the horizontally polarized components of photons $X$ and $A$ are made to interact with the coherent state using interaction parameter $3\theta$ and $-3\theta$, respectively (see Fig.~\ref{Fig:QRIO_Pic}). As a result of these interactions, the joint state evolves into the following form:
\begin{equation}
\label{Eq. JointState01}
\begin{split}
|\Phi_1\rangle & =\{\mu|Hx_0\rangle|Ha_0\rangle|Hb_0\rangle|+\eta|Vx_0\rangle|Va_0\rangle|Vb_0\rangle\\
& +\gamma|Hx_1\rangle|Ha_1\rangle|Hb_1\rangle|+\delta|Vx_1\rangle|Va_1\rangle|Vb_1\rangle\}|\alpha\rangle\\
&+\{\mu|Hx_0\rangle|Ha_1\rangle|Hb_1\rangle+\eta|Vx_0\rangle|Va_1\rangle|Vb_1\rangle\}|\alpha e^{i\theta}\rangle\\
&+\{\gamma|Hx_1\rangle|Ha_0\rangle|Hb_0\rangle+\delta|Vx_1\rangle|Va_0\rangle|Vb_0\rangle\}|\alpha e^{-i\theta}\rangle\\
&+\{\mu|Hx_0\rangle|Va_0\rangle|Vb_0\rangle+\eta|Vx_0\rangle|Ha_0\rangle|Hb_0\rangle\}|\alpha e^{3i\theta}\rangle\\
&+\{\gamma|Hx_1\rangle|Va_1\rangle|Vb_1\rangle+\delta|Vx_1\rangle|Ha_1\rangle|Hb_1\rangle\}|\alpha e^{-3i\theta}\rangle\\
&+\mu|Hx_0\rangle|Va_1\rangle|Vb_1\rangle|\alpha e^{i4\theta}\rangle\\
&+\eta|Vx_0\rangle|Ha_1\rangle|Hb_1\rangle|\alpha e^{-2i\theta}\rangle\\
&+\gamma|Hx_1\rangle|Va_0\rangle|Vb_0\rangle|\alpha e^{2i\theta}\rangle\\
&+\delta|Vx_1\rangle|Ha_0\rangle|Hb_0\rangle|\alpha e^{-4i\theta}\rangle.
\end{split}
\end{equation}
It is important to note that the states $|\alpha e^{i\theta}\rangle$ and $|\alpha e^{-i\theta}\rangle$ correspond to identical Gaussian distributions, and therefore cannot be distinguished using homodyne measurement. In contrast, the states $|\alpha\rangle$ and $|\alpha e^{\pm i\theta}\rangle$ correspond to different Gaussian distributions and can thus be distinguished \cite{PhyScr24QRIO}. Taking this observation into account, the above equation can be rewritten as
\begin{equation}
\label{Eq. JointState02}
\begin{split}
|\Phi_2\rangle & =\{\mu|Hx_0\rangle|Ha_0\rangle|Hb_0\rangle|+\eta|Vx_0\rangle|Va_0\rangle|Vb_0\rangle\\
& +\gamma|Hx_1\rangle|Ha_1\rangle|Hb_1\rangle|+\delta|Vx_1\rangle|Va_1\rangle|Vb_1\rangle\}|\alpha\rangle\\
&+\{\mu|Hx_0\rangle|Ha_1\rangle|Hb_1\rangle+\eta|Vx_0\rangle|Va_1\rangle|Vb_1\rangle\\
&+\gamma|Hx_1\rangle|Ha_0\rangle|Hb_0\rangle+\delta|Vx_1\rangle|Va_0\rangle|Vb_0\rangle\}|\alpha e^{\pm i\theta}\rangle\\
&+\{\mu|Hx_0\rangle|Va_0\rangle|Vb_0\rangle+\eta|Vx_0\rangle|Ha_0\rangle|Hb_0\rangle\\
&+\gamma|Hx_1\rangle|Va_1\rangle|Vb_1\rangle+\delta|Vx_1\rangle|Ha_1\rangle|Hb_1\rangle\}|\alpha e^{\pm 3i\theta}\rangle\\
&+\{\mu|Hx_0\rangle|Va_1\rangle|Vb_1\rangle+\delta|Vx_1\rangle|Ha_0\rangle|Hb_0\rangle\}|\alpha e^{\pm 4i\theta_1}\rangle\\
&+\{\eta|Vx_0\rangle|Ha_1\rangle|Hb_1\rangle+\gamma|Hx_1\rangle|Va_0\rangle|Vb_0\rangle\}|\alpha e^{\pm 2i\theta}\rangle.\\
\end{split}
\end{equation}
\begin{table}
\caption{Local operations performed on qubits $A$ and $B$ to obtain the entangled state given in Eq.~\eqref{Eq.Entg01.}. Here, $\sigma^s_x$ and $\sigma^p_x$ denote the Pauli-X operators acting on the spatial and polarization DOFs, respectively. The symbols Y (N) indicate whether the protocol proceeds further (does not proceed) based on the corresponding measurement outcome.}
 \label{Tab:LO_AB}
 \centering
 \begin{tabular}{p{0.38\columnwidth}p{0.22\columnwidth}p{0.22\columnwidth}c}
 \hline
 \shortstack{Measurement\\outcome\\on CS} & \shortstack{Local\\operation\\on qubit A} & \shortstack{Local\\operation\\on qubit B} & Proceed\\ [0.5ex] 
 \hline\hline
 $|\alpha\rangle$ & $-$ & $-$ & Y\\ 
 \hline
 $|\alpha e^{\pm i\theta}\rangle$ & $\sigma^s_x$ & $\sigma^s_x$ & Y\\
 \hline
 $|\alpha e^{\pm 3i\theta}\rangle$ & $\sigma^p_x$ & $\sigma^p_x$ & Y\\ 
 \hline
  $|\alpha e^{\pm 2i\theta}\rangle$ or $|\alpha e^{\pm 4i\theta}\rangle$ & $-$ & $-$ & N\\ 
 \hline
 \end{tabular}
\end{table}
Alice then performs a $X$-quadrature measurement on the CS and announces the measurement results to Bob over the classical channel. Depending on the measurement outcome, Alice and Bob apply the corresponding local operations on their qubits, as summarized in Table~\ref{Tab:LO_AB}. As a consequence of the measurement and subsequent conditional operations, the remaining photons collapse into the following entangled state:
\begin{equation}
\label{Eq.Entg01.}
\begin{split}
|\Phi_3\rangle &=\mu|Hx_0\rangle_X|Ha_0\rangle_A|Hb_0\rangle_B+\eta|Vx_0\rangle_X|Va_0\rangle_A|Vb_0\rangle_B\\
&\quad+\gamma|Hx_1\rangle_X|Ha_1\rangle_A|Hb_1\rangle_B\\
&\quad+\delta|Vx_1\rangle_X|Va_1\rangle_A|Vb_1\rangle_B.
\end{split}
\end{equation}
Once the entangled state is established, Bob applies the desired hybrid operator $U=U^P\otimes U^S$ on his qubit $B$, which transforms the joint state into the following form:
\begin{equation}
\begin{split}
|\Phi_4\rangle &=\mu^{\prime}|Hx_0\rangle_X|Ha_0\rangle_A|Hb_0\rangle_B+\eta^{\prime}|Vx_0\rangle_X|Va_0\rangle_A|Vb_0\rangle_B\\
&\quad+\gamma^{\prime}|Hx_1\rangle_X|Ha_1\rangle_A|Hb_1\rangle_B\\
&\quad+\delta^{\prime}|Vx_1\rangle_X|Va_1\rangle_A|Vb_1\rangle_B.
\end{split}
\end{equation}
Finally, Alice and Bob interfere the spatial modes of their photons $A$ and $B$ on balanced beam splitters (BBSs). Subsequently, each output path passes through a half-wave plate (HWP) oriented at $22.5^o$ followed by a polarizing beam splitter (PBS), as shown in Fig.~\ref{Fig:QRIO_Pic}. Based on the detector clicks obtained after the PBSs, Alice and Bob apply the appropriate local Pauli operations to recover the desired output state
\begin{equation} 
|\phi^{\prime}\rangle_{X} =\mu^{\prime}|Hx_0\rangle+\eta^{\prime}|Vx_0\rangle+\gamma^{\prime}|Hx_1\rangle+\delta^{\prime}|Vx_1\rangle,
\end{equation}
which is equivalent to the state obtained after the remote implementation of the hybrid operator $|\phi^{\prime}\rangle_X=(U^{P}\otimes U^{S})|\phi\rangle_X$. The correspondence between the detector outcomes and the required local Pauli corrections is summarized in Table~\ref{Tab:Final State}. Thus, the proposed protocol successfully realizes the remote implementation of an arbitrary hybrid operator on an unknown single-photon two-qubit state encoded in the polarization and spatial degrees of freedom.
\begin{table}
\caption{Local Pauli operations applied on qubits $A$ and $B$ to obtain the final desired state. Here, $\sigma^s_m$ and $\sigma^p_m$ denote the Pauli-$m$ $(m\in\{X,Y,Z\})$ operators acting on the spatial and polarization DOFs, respectively.}
 \label{Tab:Final State}
 \centering
 \begin{tabular}{m{35mm} m{35mm} m{15mm}}
 \hline
 Measurement outcome on A & Measurement outcome on B & Operation \\ [0.5ex] 
 \hline\hline
 $|Ha_0\rangle$ & $|Hb_0\rangle$ & $I$\\ 
 \hline
 $|Ha_0\rangle$ & $|Hb_1\rangle$ & $\sigma_z^s$\\
 \hline
 $|Ha_1\rangle$ & $|Hb_0\rangle$ & $\sigma_z^s$\\
 \hline
 $|Ha_1\rangle$ & $|Hb_1\rangle$ & $I$\\
 \hline
 $|Ha_0\rangle$ & $|Vb_0\rangle$ & $\sigma_z^p$\\ 
 \hline
 $|Ha_0\rangle$ & $|Vb_1\rangle$ & $\sigma_z^s\sigma_z^p$\\ 
 \hline
 $|Ha_1\rangle$ & $|Vb_0\rangle$ & $\sigma_z^s\sigma_z^p$\\ 
 \hline
 $|Ha_1\rangle$ & $|Vb_1\rangle$ & $\sigma_z^p$\\ 
 \hline
 $|Va_0\rangle$ & $|Hb_0\rangle$ & $\sigma_z^p$\\ 
 \hline
 $|Va_0\rangle$ & $|Hb_1\rangle$ & $\sigma_z^s\sigma_z^p$\\ 
 \hline
 $|Va_1\rangle$ & $|Hb_0\rangle$ & $\sigma_z^s\sigma_z^p$\\ 
 \hline
 $|Va_1\rangle$ & $|Hb_1\rangle$ & $\sigma_z^p$\\ 
 \hline
 $|Va_0\rangle$ & $|Vb_0\rangle$ & $I$\\ 
 \hline
 $|Va_0\rangle$ & $|Vb_1\rangle$ & $\sigma_z^s$\\ 
 \hline
 $|Va_1\rangle$ & $|Vb_0\rangle$ & $\sigma_z^s$\\ 
 \hline
 $|Va_1\rangle$ & $|Vb_1\rangle$ & $I$\\ 
 \hline
 \end{tabular}
\end{table}
\section{Results and Discussion}\label{Sec:results}
The performance of the protocol described in Sec.~\ref{Sec:qrio_protocol} is primarily determined by the outcomes of the homodyne measurement performed on the auxiliary CS. In the ideal case, the measurement can yield five distinguishable outcomes, namely $|\alpha\rangle,\,|\alpha e^{\pm i\theta}\rangle,\,|\alpha e^{\pm i2\theta}\rangle,\,|\alpha e^{\pm i3\theta}\rangle,\,\text{and}\,|\alpha e^{\pm i4\theta}\rangle.$ Among these five possible outcomes, only three correspond to successful implementation of the protocol (see Table~\ref{Tab:LO_AB}). Therefore, the ideal success probability of the protocol is $P_{\mathrm{succ}}=\frac{3}{5}$. However, the finite overlap between neighboring coherent states introduces measurement errors. The inability to perfectly distinguish the pairs of states $|\alpha\rangle$ and $|\alpha e^{\pm i\theta}\rangle$, $|\alpha e^{\pm i\theta}\rangle$ and $|\alpha e^{\pm i2\theta}\rangle$, $|\alpha e^{\pm i2\theta}\rangle$ and $|\alpha e^{\pm i3\theta}\rangle$, and $|\alpha e^{\pm i3\theta}\rangle$ and $|\alpha e^{\pm i4\theta}\rangle$ gives rise to the corresponding error probabilities
\begin{align*}   
    p_{\mathrm{err1}} &= \frac{1}{2}\mathrm{erfc}\!\left[\frac{\alpha(1-\cos\theta)}{ \sqrt{2}}\right],\\
    p_{\mathrm{err2}} &= \frac{1}{2}\mathrm{erfc}\!\left[\frac{\alpha(\cos\theta-\cos2\theta)}{ \sqrt{2}}\right],\\
    p_{\mathrm{err3}} &= \frac{1}{2}\mathrm{erfc}\!\left[\frac{\alpha(\cos2\theta-\cos3\theta)}{ \sqrt{2}}\right],\\
    p_{\mathrm{err4}} &= \frac{1}{2}\mathrm{erfc}\!\left[\frac{\alpha(\cos3\theta-\cos4\theta)}{\sqrt{2}}\right],
\end{align*}
respectively. Furthermore, the coherent state undergoes photon loss due to its interaction with the environment during the cK-interaction and the subsequent homodyne measurement. This dissipation effectively reduces the amplitude of the coherent state, thereby increasing the measurement error. Accounting for this effect, the error probabilities become
\begin{align*}   
    p_{\mathrm{err1}} &= \frac{1}{2}\mathrm{erfc}\!\left[\frac{D\alpha(1-\cos\theta)}{ \sqrt{2}}\right],\\
    p_{\mathrm{err2}} &= \frac{1}{2}\mathrm{erfc}\!\left[\frac{D\alpha(\cos\theta-\cos2\theta)}{ \sqrt{2}}\right],\\
    p_{\mathrm{err3}} &= \frac{1}{2}\mathrm{erfc}\!\left[\frac{D\alpha(\cos2\theta-\cos3\theta)}{ \sqrt{2}}\right],\\
    p_{\mathrm{err4}} &= \frac{1}{2}\mathrm{erfc}\!\left[\frac{D\alpha(\cos3\theta-\cos4\theta)}{\sqrt{2}}\right],
\end{align*}
where $D=e^{-\gamma t}$ with $\gamma$ denoting the dissipation rate and $t$ representing the interaction time~\cite{PhyScr24QRIO}. Taking into account both the finite distinguishability of the CS and the dissipation-induced reduction in the coherent state amplitude, the success probability of the proposed protocol is modified to
\begin{equation}
\label{eq:P_succ}
P_{\mathrm{succ}}=\frac{3}{5}\left(1-p_{\mathrm{err1}}-p_{\mathrm{err2}}-p_{\mathrm{err3}}-p_{\mathrm{err4}}\right).    
\end{equation}
Above equation is obtained by considering that the error events are mutually exclusive. The rationale behind this intrinsic assumption follows from Eq.~\eqref{Eq. JointState02}, where we can easily see that the measurement outcomes $|\alpha\rangle\,,|\alpha e^{\pm i\theta}\rangle\,,|\alpha e^{\pm i2\theta}\rangle\,,|\alpha e^{\pm i3\theta}\rangle$ are mutually exclusive. This expression shows that the protocol success probability is reduced from its ideal value of $3/5$ due to the combined effects of measurement errors and coherent state dissipation.

\begin{figure}
    \centering
    \includegraphics[width=0.8\linewidth]{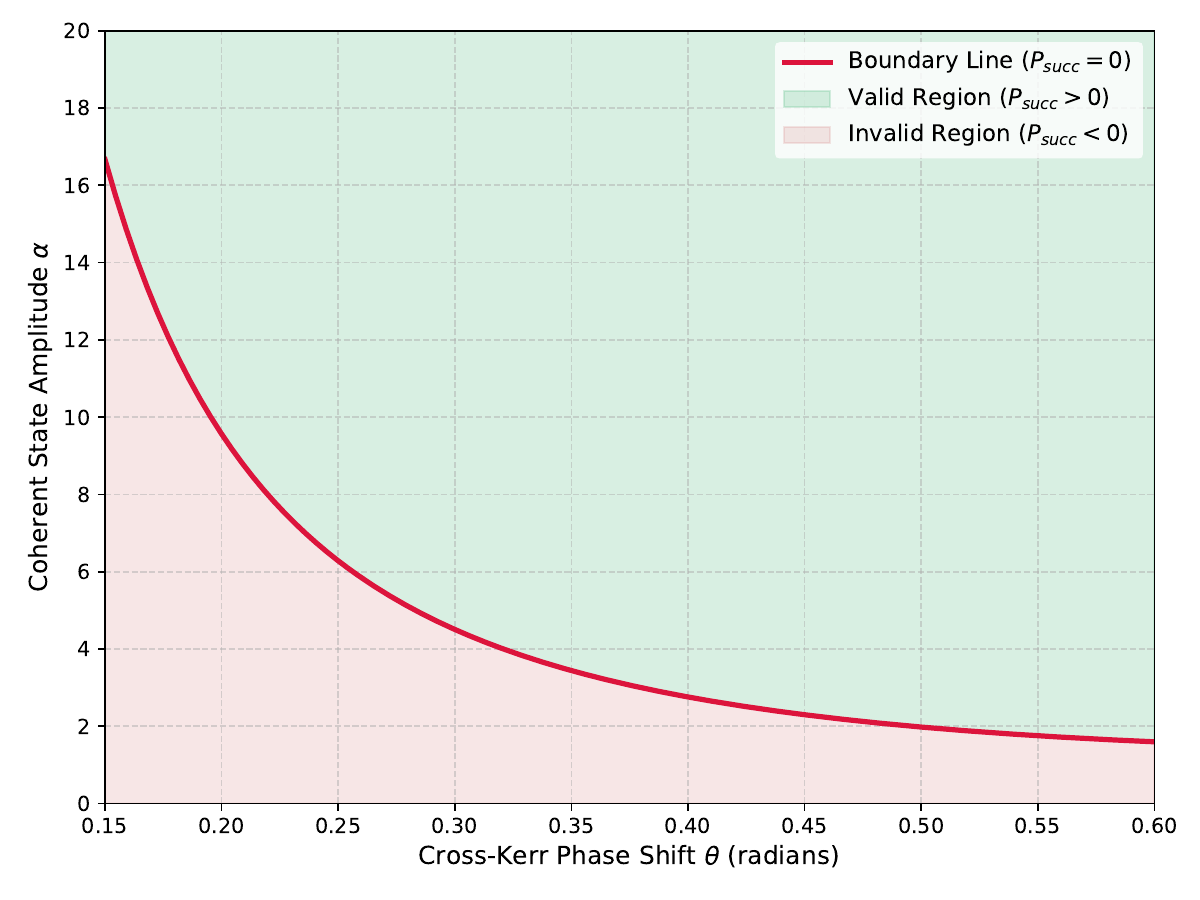}
    \caption{(Color online) Parameter region for positive success probability of the proposed QRIO protocol in the $(\theta,\alpha)$ plane. The boundary curve $(P_{\mathrm{succ}}=0)$ separates the valid region ($(P_{\mathrm{succ}}>0)$, shaded green) from the invalid region ($(P_{\mathrm{succ}}<0)$, shaded red). For a given cross-Kerr phase shift $\theta$, the coherent state amplitude $\alpha$ must lie above the boundary to ensure a positive success probability.}
    \label{fig:valid2dplot}
\end{figure}

Given the expression for the success probability in Eq.~\eqref{eq:P_succ}, it is first necessary to determine the range of the cross-Kerr phase shift $\theta$ and the CS amplitude $\alpha$ for which the protocol yields a physically meaningful success probability. Since the prefactor (3/5) is always positive, the positivity of $P_{\mathrm{succ}}$ is ensured by the condition $1-p_{\mathrm{err1}}-p_{\mathrm{err2}}-p_{\mathrm{err3}}-p_{\mathrm{err4}}>0.$ Fig.~\ref{fig:valid2dplot} illustrates the boundary separating the valid and invalid parameter regions in the $(\theta,\alpha)$ plane. The boundary curve corresponds to $P_{\mathrm{succ}}=0$. The region above the boundary represents the set of parameter values for which the protocol achieves a positive success probability, whereas the region below the boundary corresponds to unphysical parameter values yielding $P_{\mathrm{succ}}<0$. The coherent state amplitude is observed to drop asymptotically with the increase in cross-Kerr nonlinearity, indicating that larger nonlinear phase shifts enhance the distinguishability of the coherent states and thereby improve the protocol performance. Between $0.3\,\mathrm{rad} \lesssim \theta \lesssim 0.6\,\mathrm{rad}$, the boundary becomes nearly flat, indicating that a relatively small coherent state amplitude, approximately $1.5 \lesssim \alpha \lesssim 4.5$ is sufficient to ensure $P_{\mathrm{succ}}>0$.

\begin{figure*}[t]
    \centering
    \subfloat[]{%
    \includegraphics[width=0.48\textwidth]{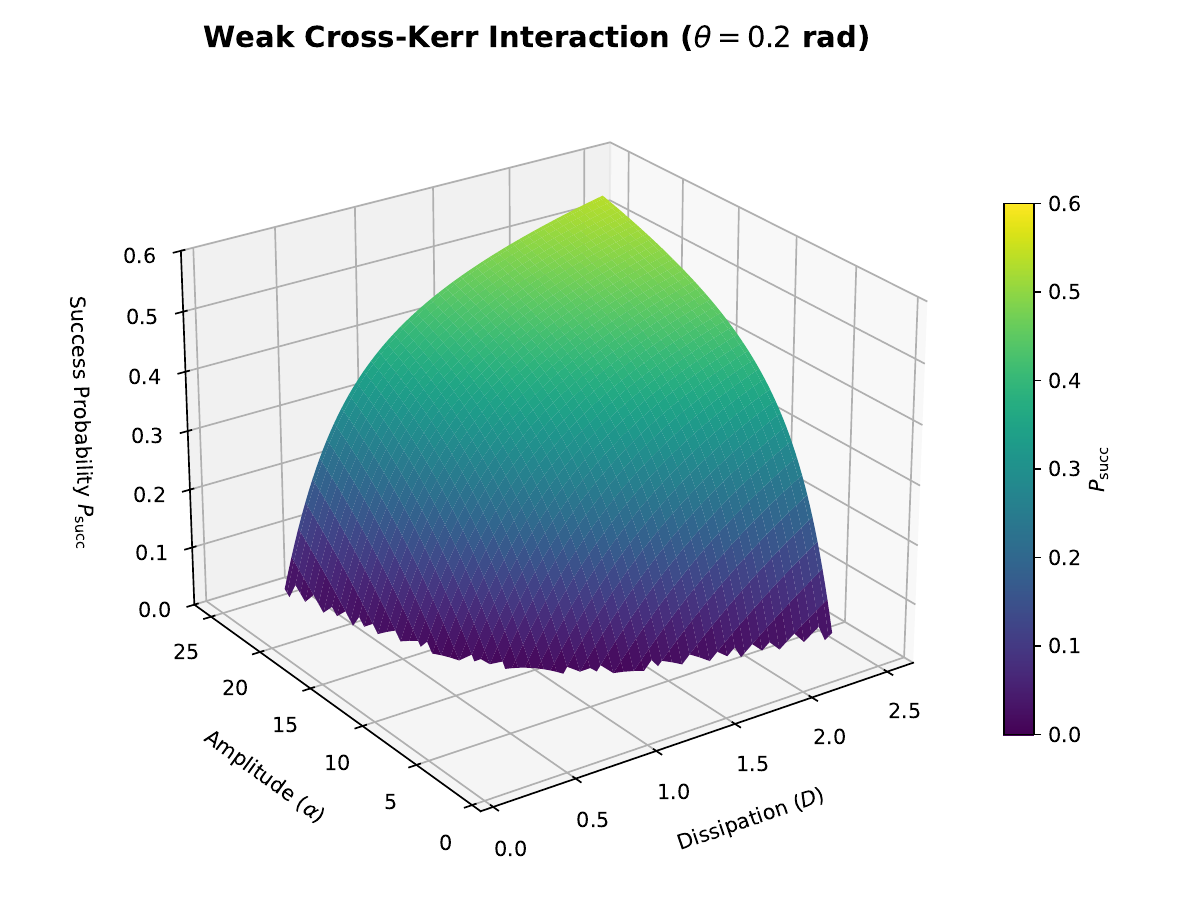}}
    \label{fig:3d1}
    \hfill
    \subfloat[]{%
    \includegraphics[width=0.48\textwidth]{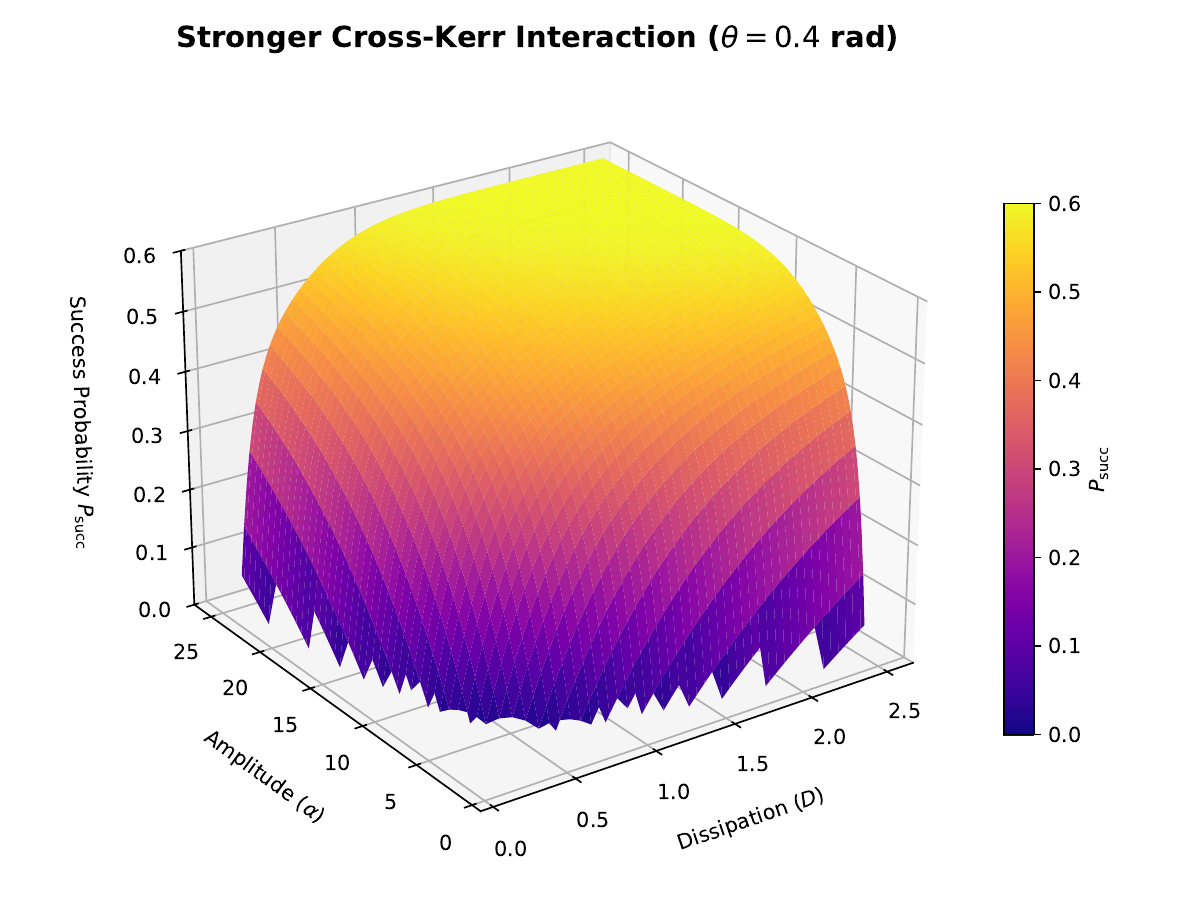}}
    \label{fig:3d2}
    \vspace{0.5cm}
    \subfloat[]{%
    \includegraphics[width=0.8\textwidth]{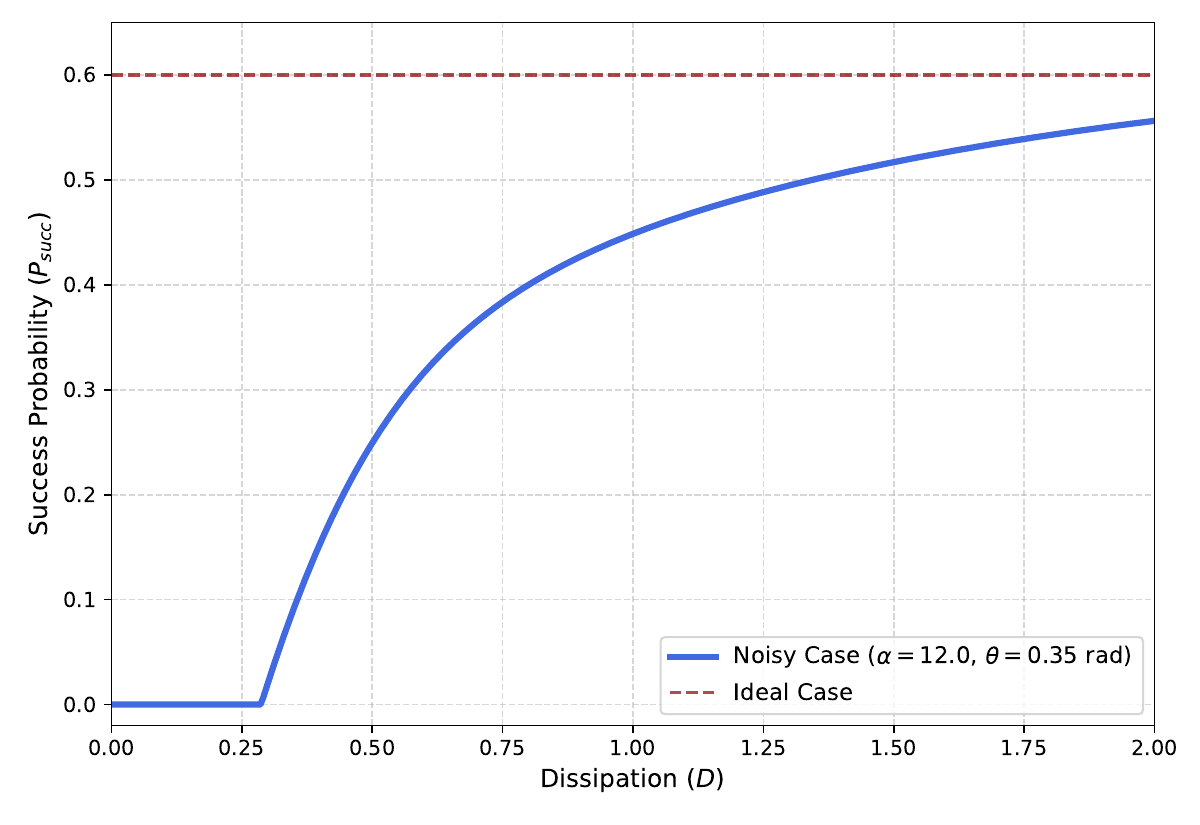}}    
    \label{fig:3d3}
    \caption{(Color online) Effect of coherent state dissipation on the success probability of the proposed QRIO protocol. (a) Three-dimensional variation of the success probability $P_{\mathrm{succ}}$ as a function of the dissipation parameter $D$ and coherent state amplitude $\alpha$ for a weak cross-Kerr phase shift $(\theta=0.2$ rad). (b) Corresponding success probability landscape for a stronger cross-Kerr phase shift $(\theta=0.4$ rad). In both cases, larger values of $\alpha$ and weaker dissipation $(D\rightarrow1)$ enhance the protocol performance. (c) Comparison of the ideal $(D=1)$ and noisy success probabilities as a function of the dissipation parameter for $\alpha=12$ and $\theta=0.35$ rad, illustrating the degradation in performance caused by coherent state dissipation.}
    \label{fig:dissipation_plot}
\end{figure*}

Having identified the allowed values of the cross-Kerr phase shift $\theta$ and the coherent state amplitude $\alpha$, we next investigate the influence of the coherent state dissipation parameter $D$ on the success probability of the proposed protocol. Fig.~\ref{fig:dissipation_plot} presents the variation of $P_{\mathrm{succ}}$ with $D$ and $\alpha$ for two representative values of the cross-Kerr phase shift, namely $\theta=0.2$ rad and $\theta=0.4$ rad together with a two-dimensional comparison of the ideal and noisy cases. Fig.~\ref{fig:dissipation_plot}(a) and (b) show the three-dimensional success probability landscapes for weak and relatively stronger cross-Kerr nonlinearities, respectively. In both cases, the success probability increases monotonically with the coherent state amplitude $\alpha$ and the dissipation parameter $D$. The highest success probability is achieved in the regime of large $\alpha$ and $D\approx1$, corresponding to negligible coherent state dissipation. A comparison of Fig.~\ref{fig:dissipation_plot}(a) and (b) further reveals that increasing the cross-Kerr phase shift significantly enhances the protocol performance. For $\theta=0.4$ rad, the protocol attains a high success probability even for relatively small coherent state amplitudes, whereas a much larger amplitude is required when $\theta=0.2$ rad. In other words, when the cross-Kerr parameter is stronger, the protocol activates at much lower dissipation values $D$ and low laser powers $\alpha$. This improvement arises from the increased separation between the coherent state distributions which reduces the measurement error during homodyne detection. To further illustrate the effect of dissipation, Fig.~\ref{fig:dissipation_plot}(c) compares the success probability in the ideal $(D=1)$ and noisy cases for fixed values of $\alpha=12$ and $\theta=0.35$ rad. The ideal success probability remains constant at $3/5$, whereas the noisy success probability decreases monotonically with increasing dissipation. As dissipation drops (increasing $D$), the shifted coherent states separate cleanly and the success profile rapidly climbs. At low values of $D$, dissipation completely ruins state distinguishability and the success probability sits firmly at zeros. These results demonstrate that minimizing coherent state dissipation while optimizing the cross-Kerr phase shift $\theta$ and coherent state amplitude $\alpha$ significantly improves the success probability of the proposed protocol for the remote implementation of a hybrid operator.
\textcolor{black}{It is naturally expected that similar situations will exist for the other protocols involving similar resources, and the approach adopted here will be of much use for analyzing the meaningfulness of existing and futuristic work in this domain.}

\section{Conclusion}\label{Sec:conclusion}
In this work, we have proposed a protocol for the quantum remote implementation of an arbitrary hybrid operator acting on a single-photon two-qubit hybrid state encoded in the polarization and spatial DOFs. Unlike existing QRIO schemes that employ hyperentangled states while utilizing only one degree of freedom at a time, the proposed protocol simultaneously exploits both the polarization and spatial DOFs of a polarization-path entangled state. This simultaneous utilization enables the direct remote implementation of a hybrid operator of the form $U^{P}\otimes U^{S}$, thereby providing a more resource efficient approach for hybrid quantum information processing.

The protocol is constructed using linear optical components, cross-Kerr nonlinear interactions and homodyne measurements, making it compatible with the current toolbox of optical quantum technologies. We have presented the complete sequence of operations required for entangling the unknown hybrid state with the shared entangled resource, performing the remote implementation and recovering the desired output state through appropriate local Pauli corrections.

To evaluate the practicality of the proposed scheme, we have analyzed the effects of two important experimental imperfections: finite distinguishability of coherent states during homodyne detection and coherent state dissipation. Analytical expressions for the corresponding measurement error probabilities were derived, leading to a modified expression for the protocol success probability. We further identified the permissible parameter region of the cross-Kerr phase shift and coherent state amplitude that guarantees a physically meaningful success probability. The numerical results show that increasing the cross-Kerr phase shift enhances the distinguishability of the coherent states, thereby reducing the measurement error and allowing the protocol to operate with smaller coherent state amplitudes. Furthermore, minimizing coherent state dissipation significantly improves the protocol performance, enabling success probabilities close to the ideal value.

The proposed protocol provides a feasible approach for the remote implementation of hybrid quantum operations and may find applications in distributed quantum computation, quantum communication networks and future hybrid photonic quantum information processing architectures. We believe that the present work opens new avenues for exploiting multiple DOFs simultaneously in remote quantum information processing tasks and motivates further investigations toward experimentally realizable hybrid quantum networks.

\section*{Acknowledgment}
The authors thank the Department of Science and Technology, Government of India for the support provided through the National Quantum Mission (NQM). 

\section*{Competing Interests}
Authors declare that they don't have any competing interests.

\section*{Data Availability}
No additional data is generated through this work. All the relevant data is included in the paper.



\bibliography{apssamp}

\end{document}